\documentclass[aps,prl,floatfix,twocolumn,showpacs]{revtex4}
\usepackage{graphicx} \usepackage{amsmath} \usepackage{amssymb}
\newcommand{\BEQ}{\begin{equation}}
\newcommand{\EEQ}{\end{equation}}
\newcommand{\BEA}{\begin{eqnarray}}
\newcommand{\EEA}{\end{eqnarray}}

\renewcommand{\d}{{\rm d}}

\begin{document}
\draft
\title
{Adiabatic feedback
control of Hamiltonian systems.}
\date{\today}
\author{A.E. Allahverdyan, K.G. Petrosyan and D.B. Saakian}
\address{Yerevan Physics Institute,
Alikhanian Brothers St. 2, Yerevan 375036, Armenia.}

\begin{abstract} We study feedback control of classical Hamiltonian
systems with the controlling parameter varying slowly in time.  The control aims
to change system's energy.  We show that the control problems
can be solved with help of an adiabatic invariant that generalizes the
conservation of the phase-space volume to control situations.  New
mechanisms of control for achieving heating, cooling, entropy reduction
and particle trapping are found.  The feedback control of a many-body
system via one of its coordinates is discussed.  The results are
illustrated by two basic models of non-linear physics. 

\end{abstract}






\maketitle

Physics met control theory yet in XIX'th century: the notorious
Maxwell's demon poses a problem of building a control system that will
reduce the entropy (increase the order) of a statistical system
\cite{demon,pop}. Founders of cybernetics recognized the entropy
reduction as one of the basic goals of control \cite{cyber,cyber_ash}.
This is especially relevant, since the statistical description|in
particular, thermodynamical relations and concepts such as entropy| are
needed not only for macroscopic systems, but also for few-body chaotic
ones \cite{zas,berdi}.  Deep implications of the
entropy-information-control relations were studied in
\cite{pop,cyber_ash}. Theoretical and experimental methods of
controlling physical systems have recently undergone an explosive
development \cite{bech,fradkov,rev,exp}. The focus is on non-linear and
chaotic systems in view of their numerous applications. 

The potential of physical ideas applicable to the control science is,
however, far from being exhausted.  Here we explore a feedback control
of ergodic Hamiltonian systems, where a slow motion of the controlling
parameter leads to the existence of an adiabatic invariant.  This fact
helps to solve the control problem on a general level.  Feedback means
that the control parameter depends on the current state of the system,
while in the non-feedback control the motion of the control parameter is
prescribed.  In this respect we built up on the non-feedback adiabatic
control of ergodic systems, where the adiabatic invariance of the
phase-space volume is well-known \cite{hertz,berdi,ott}.  Despite of its
general importance, from the control viewpoint this method has several
drawbacks: {\it i)} The microcanonic entropy is conserved (or increases
for non-adiabatic processes \cite{ott}), so that the goal of the control
is not achieved. {\it ii)} The method does not function for the
important case of cyclic influences of the controlling parameters, i.e.,
it insists on permanent modifications of the system.  We show that these
drawbacks are absent for adiabatic feedback control, while the main
advantage|independence on the details of the system under assuming
ergodicity of certain observables|is kept. 

Consider a system with $N$ degrees of freedom and Hamiltonian
$H(p,q,R)$. The equations of motion read \BEA \label{1}
\dot{p}=-\partial_q H(p,q,R),\qquad \dot{q}=\partial_p H(p,q,R),
\EEA where $q=(q_1,...,q_N)$ and $p=(p_1,...,p_N)$ are,
respectively, the coordinates and momenta, and where $R$ is the
control parameter. The purpose of varying $R$
in time is to change system's energy $E$ in a desired way. This
setup of control \cite{bech,fradkov} is especially relevant
for chaotic systems, since energy is the basic constant of motion
\cite{zas,berdi}.  

The change of $R$ can be sizable, but the
speed of $R$ is small and depends on the dynamical variables $p$
and $q$ via a phase-space observable $F$ (feedback). Behind the scene of this
description is a measurement of $F$ (sensor) and subsequent engineering (actuator)
that leads to 
\BEA \label{2} 
\tau_R\dot{R}=F(z,R), \quad z\equiv(q,p),
\EEA where
$\tau_R$ is much larger than the
characteristic time $\tau_S$ of the
system (defined with respect to the $R=$const 
dynamics).  The derivative of the control parameter depending on the
state of the system is standard in control practice
\cite{bech,fradkov}. For a general $F$ the problem is
hardly tractable analytically. However, an important paper
\cite{lisak} shows that an one-dimensional ($N=1$) Hamiltonian system admits
an adiabatic invariant provided 
\BEA \label{3}
F(z,R)=f(R)\,\varphi(z). 
\EEA 
Here $\varphi$ and $f$ depend on
the fast and slow variables, respectively; these dependencies are
thus factorized in (\ref{3}). Assuming (\ref{3}), we extend this
invariant to systems with ergodic observables and apply it to control
problems.

The below derivation is based on the fact
that $R$ and energy $E$ are slow variables, since they change on times $\sim\tau_R$.
For times $\tau_R\gg\tau\gg\tau_S$ we have from (\ref{1}, \ref{2}, \ref{3}):
\begin{gather}
\frac{\d E}{\d \tau}\equiv
\frac{1}{\tau}\,[\,H(z_{t+\tau},R_{t+\tau})-H(z_t,R_t)\,]\\=
\int_t^{t+\tau}\frac{\d s}{\tau}\,\frac{\d H}{\d s}(z_s,R_s)
=\int_t^{t+\tau}\frac{\d s}{\tau}\,\dot{R}_s\,\frac{\partial H}{\partial R}(z_s,R_s),
\nonumber\\
=\frac{1}{\tau_R}\,f(R_t)
\int_t^{t+\tau}\frac{\d s}{\tau}\,
\varphi(z)\,\frac{\partial H}{\partial R}(z_s,R_t)
+o(\frac{\tau}{\tau_R}).
\label{hek}
\end{gather}
The last integral refers to the dynamics with $R_t=$const.
Now note the Liouville theorem
$\d z=\d z_t$ and energy conservation $H(z_{t+\tau})=H(z_t)=E_t$,
and denote $w(z)\equiv\varphi(z)\,\partial_RH(z,R_t)$. Noting
the microcanonical distribution
\BEA
{\cal M}(z)=\frac{\delta[E-H(z,R)]}{\partial_E\Omega(E,R)},~~
\Omega(E)=\int\d z\, \theta[E-H(z,R)],\nonumber
\label{quito}
\EEA
where $\delta(x)$ and $\theta(x)$ are, respectively, the delta and step function, and 
where $\Omega$ is the phase-space volume, we get
\BEA
\label{karamba1}
\int \d z \,w(z){\cal M}(z,E_t)
=\frac{1}{\tau}\int_t^{t+\tau}\d s
\int \d z \,w(z){\cal M}(z,E_t)\\
=\int \d z_t\,{\cal M}(z_t,E_t)\,
\frac{1}{\tau}\int_t^{t+\tau}\d s \,w(z[z_t,s]),
\label{karamba2}
\EEA
where $z[z_t,s]$ is the trajectory at time $s$ with the initial condition $z_t$.
If $w(z,R)$ is an {\it ergodic
observable} of the $R_t=$const dynamics, then
for $\tau\gg \tau_S$ the time-average in (\ref{karamba2}) does not
depend on the initial condition $z_t$ \cite{berdi}, the integration over
$z_t$ in (\ref{karamba2}) drops out, and we get from (\ref{karamba1})
that the time-average in (\ref{hek}) is equal to the
microcanonical average at the energy $E_t$:
\BEA
\frac{\d E}{\d \tau}
=\frac{f(R_t)}{\tau_R}
\int \d z\, {\cal M}(z,E_t,R_t)
\varphi(z)\,\frac{\partial H}{\partial R}(z_s,R_t).
\label{hek1}
\EEA
Assuming ergodicity of $\varphi(z)$ we get from (\ref{2}, \ref{3})
\BEA
\frac{\d R}{\d \tau}
=\frac{1}{\tau_R}\,f(R_t)
\int \d z\, {\cal M}(z,E_t,R_t)
\varphi(z).
\label{hek2}
\EEA
It is seen that (\ref{hek1}, \ref{hek2}) amount to
the conservation condition
\BEA
\frac{\d I}{\d \tau}=0,~~ I(E,R)\equiv\int \d z\, \varphi(z)\theta[E-H(z,R)].
\label{ff}
\EEA
$I$ does not depend on the initial condition $z(t_{\rm i})$ of the
trajectory. Given the initial values $E_{\rm i}$ and $R_{\rm i}$, the final
$E_{\rm f}$ and $R_{\rm f}$ are found {\it self-consistently} from (\ref{hek2})
and from
\BEA
\label{de}
I(E_{\rm i}, R_{\rm i})=
I(E_{\rm f}, R_{\rm f}).
\EEA
For $\varphi=1$ (no feedback) we get from (\ref{ff}) the conservation of
the phase-space volume $\Omega$. For an isolated ergodic system the
entropy is defined as $S=\ln \Omega$ \cite{berdi}. This definition
satisfies to all reasonable features of entropy, e.g., for the
temperature defined via $1/T=\partial_E\ln \Omega(E)$, the integration
by parts leads to equipartition \cite{berdi}: $\langle x\,\partial_x
H\rangle=T$, where $x$ is any canonical coordinate of momentum, and
$\langle ...\rangle$ is the average over microcanonical distribution
(\ref{quito}).  For non-adiabatic processes $S$ normally increases, thus
confirming the second law \cite{berdi,ott}. In the standard
thermodynamical limit $S$ goes to the more usual expression
$\widetilde{S}=\ln [\partial_E\Omega]$ \cite{berdi}. {\it Neither} of the
above important features holds if we apply $\widetilde{S}$ for a finite
system \cite{berdi}.

For the feedback case $\varphi\not=1$, (\ref{de}) shows that $\Omega$ is
{\it not} conserved, and the entropy $S$ can decrease, as we see below.
Thus the basic goal of control will be attained.

As an application consider $n$ particles moving between an
infinite wall located at $q=0$ and a piston with the controlled
coordinate $q=R>0$.  The Hamiltonian is
$H=\sum_{i=1}^n\frac{p_i^2}{2m_i}$ ($m_i$ are masses) apart from the
infinite wall interactions that lead to elastic scattering and apart from
weak inter-particle couplings that ensure the ergodicity of the
system \cite{berdi}. The simplest type of feedback is 
realized via the coordinate $q$ of one of the
particles: $\varphi=\varphi(q)$. The conservation of $I$ leads from (\ref{ff})
\BEA \label{tartar} E_{\rm i}^{n/2}\int_0^{R_{\rm i}}\d q\,
\varphi(q)\, q^{n-1}=E_{\rm f}^{n/2}\int_0^{R_{\rm f}}\d q\,
\varphi(q)\, q^{n-1}.
\EEA We start with $n=1$ (Fermi accelerator), where
the $R=$const motion is obviously ergodic. 
This model was intensively studied in plasma and accelerator
physics, nonlinear physics and astrophysics \cite{zas,liebe_g}. A
question that created the interest to the model is how to increase
system's energy $E$ (heating)?  The non-feedback adiabatic control
is able to heat up the system only upon decreasing $R$, since the
phase-space volume $\propto\sqrt{E}R$ is conserved.  This is not
realistic, since involves a permanent and substantial
modification of system's parameter.  A non-adiabatic heating studied
in \cite{zas,liebe_g} amounts to an oscillating motion of the
piston, e.g., $R(t)=R_{\rm i}+\alpha T\sin\frac{t}{T}$ with period
$T$.  The maximal achievable energy is $E_{\rm max}\sim R_{\rm
i}\alpha/T$ \cite{zas,liebe_g}.  Once $R_{\rm i}$ is fixed, 
$E_{\rm max}$ grows by increasing $\alpha$ (large amplitude)
or by decreasing $T$ (large piston velocity).  Thus it
is impossible to increase the energy as much as desired \cite{com1}.
The adiabatic feedback control solves this problem as follows.
Take $\varphi(q)=q-a$, where $a>0$ is a
constant. We get from (\ref{tartar})
\BEA
\label{campanero}
\sqrt{E_{\rm f}}R_{\rm f}(R_{\rm f}-2a)=
\sqrt{E_{\rm i}} R_{\rm i}(R_{\rm i}-2a).
\EEA
Taking $R_{\rm f}\to 2a$ (provided $R_{\rm i}$ is not close to
$2a$), the final energy is increased as much as desired.  Eq.~(\ref{hek2})
for the piston is $\tau_R\frac{\d}{\d \tau}
R=\frac{f}{2}(R-2a)$, or (taking $f=$const)
\BEA
\label{doka}
R(\tau)=2a+(R_{\rm i}-2a) e^{f\tau/(2\tau_R)}.
\EEA
For $f<0$, $|f\tau|\gg \tau_R$ suffices to make $R_{\rm f}$ very
close to $2a$.  Eq.~(\ref{tartar}) also shows that for a more general
feedbacks function $\varphi$, the strong heating exists for
$\int_0^{R_{\rm f}}\d q\, \varphi(q)\to 0$. 

There are three regimes of (\ref{campanero}) considered separately: {\it
i)} for $R_{\rm i}+R_{\rm f}>2a>R_{\rm f}>R_{\rm i}$ the heating is
achieved by expanding the system.  Taking the feedback out and squeezing
$R$ adiabatically back to $R_{\rm i}$, we heat up the particle even more and
complete the cycle. The reason of the feedback heating is seen from
(\ref{2}, \ref{3}): $\dot{R}=|f|(a-q)$. When $R>a$, the piston tends to
move towards the approaching particle, and during the resulting
collision the particle gains energy. However, $\dot{R}>0$ when the
particle is far from the piston, and thus in average $R$ increases. {\it
ii)} For $R_{\rm i}>R_{\rm f}>2a$ the particle is heated by squeezing.
{\it iii)} The case $R_{\rm f}<R_{\rm i}<2a$ and $R_{\rm i}+R_{\rm f}<2
a$ is interesting, since in contrast to the above two cases the heating
is not strong, but leads to the entropy {\it decrease}. Now in (\ref{doka}) we
should take $f>0$. 

The ability of energy decreasing (cooling) is not less important in
applications \cite{exp}.  A scenario of cooling is illustrated by
(\ref{campanero}) in the regime $R_{\rm i}+R_{\rm f}>2a>R_{\rm i}>R_{\rm
f}$, which is realized by taking $f>0$ in (\ref{doka}).  Now
$\dot{R}=f(q-a)$, and moving out of the approaching particle, the piston
gains energy. In the second step we take the feedback out and expand the
piston back to its original position $R_{\rm i}$, thereby cooling the
system even more, reducing its entropy and completing the cycle.  To
lower the energy as much as desired we need several cycles. This
contrasts the heating situation, where one cycle sufficed.
This is because $\int_0^{R}\d q\, \varphi(q)$ can turn to zero,
but cannot be infinite, since $\varphi$ gives the speed of
$R$.  Note that without feedback there is no obvious cooling cyclic
process. 

For a finite number of particles $n$ one can design from (\ref{tartar})
cooling and heating processes analogously to the $n=1$ case. Extensions
to more general models of chaotic systems \cite{cont},
e.g., billiards \cite{zas} are straightforward. 

What about controlling a many-body ergodic system, where $n\gg 1$ and
the feedback still goes over one coordinate: $\varphi=\varphi(q)$?  This
coordinate may belong to the particle with a large mass that moves
slowly (due to equipartition) and is amenable for the measurement. Now
note that $\varphi(q)\sim e^{bn q}$ ($|b|\sim 1$) is {\it unphysical}: if
$b>0$ this will lead to a huge speed of $L$ on the short times; see
(\ref{2}, \ref{3}). If $b<0$ this speed will be too small and the
control takes a huge time. Then for $n\gg 1$ and $R>1$ the integral in
(\ref{tartar}) is dominated by the right end: $I\simeq \varphi(R)E^{n/2}
R^{n} $.  This brings for the change of the microcanonical entropy
$S=n\ln[R\sqrt{E}]$: $S_{\rm f}-S_{\rm i}= \ln\frac{\varphi(R_{\rm
i})}{\varphi(R_{\rm f})}$.  Normally $S_{\rm f}-S_{\rm i}\sim 1$,
though with a precise tuning we can achieve $S_{\rm f}-S_{\rm i}\sim \ln n$;
take $\varphi(R)=R-a$ and $R_{\rm f}-a\sim n^{-1}$. Thus when controlling a
macroscopic system via a single degree of freedom, the effect of
feedback is small.  This conclusion is more general: if the feedback
uses the coordinate $q$, (\ref{ff}) implies $I=\int \d q\,\varphi(q)
e^{S(q)}$, where $S(q)=\ln \Omega(q)$ is the entropy of the system with
a fixed value of $q$. Once normally $S(q)\sim n$ (for $n\gg 1$), 
the conclusion follows along the above line.

Our last example is interesting in two respects. First this is a
paradigmatic model of non-linear science \cite{zas} and control 
theory \cite{fradkov}: the oscillator
$H=\frac{p^2}{2}-R\cos q$, where the amplitude $R>0$ is taken as the
control variable.  The periodic boundary conditions are ensured by
$-\pi\leq q\leq \pi$ and $\varphi(q,p)=\varphi(q+2\pi,p)$.  The model is
basic for particle-wave interactions, where the potential $-R\cos q$
models a harmonic wave (in the system of reference where the wave is
standing), and where $E>R$ and $E<R$ refer to the free (rotating) and
captured (oscillating) particles, respectively \cite{zas,liebe_g,best}.
Second, the system is non-ergodic: for $E>R$ the phase-space of the
$R=$const dynamics consists of two ergodic components supporting the
rotational motion with, respectively, $p>0$ and $p<0$ (the rotating
particle never changes the sign of its momentum, while the energy is
insensitive to this sign).  In contrast, the oscillatory motion for
$R>E>-R$ is globally ergodic.  When $R$ is time-dependent some
trajectories cross the separatrix $E=R$ and move from one ergodic
component to another. Although the model is non-ergodic, $I$ defined in
(\ref{ff}) (with the integration over the whole phase-space
irrespectively of ergodic components) is conserved if the observable
$w(z)=\varphi(q,p)\partial_RH(q,p)=-\varphi(q,p)\cos q$ is $p\to -p$
symmetric. Indeed, if two ergodic components are possible for a given
energy, $w(z)$ does not depend on which component the paticle moves.  The 
adiabatic condition
$\tau_R\gg\tau_S$ can be satisfied as well: though the $R=$const motion
on the separatrix has an infinite period (due to unstable fixed points
$q=\pm \pi$), the fraction of particles trapped by the separatrix is
negligible (measure zero) \cite{best}, so that $\tau_S$ is finite.  Then
the derivation (\ref{karamba1}--\ref{hek2}) applies. For non-feedback processes
the conservation of the phase-space volume $\Omega$ was first shown in \cite{best}.

\begin{figure}[bhb]
\hfill
\includegraphics[width=8.8cm]{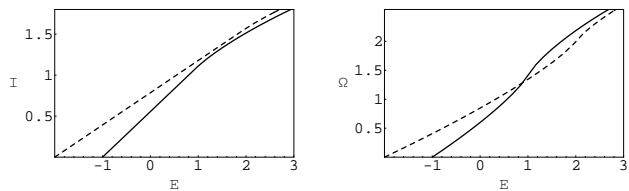}
\caption{The adiabatic invariant $I$ and the phase-space volume $\Omega$ versus energy
$E$. Dotted lines $R=2$. Normal lines $R=1$.}
\hfill
\label{fig_1}
\end{figure}
We now focus on the following question: how to trap
particles by means of a cyclic change of the amplitude $R$? This is
a version of the cooling problem. First note that the phase-space
volume $\Omega$ is expressed from (\ref{karamba1}) as (changing
variables for (\ref{ga1}) as $\epsilon\sin\xi=\sin\frac{q}{2}$) \BEA
\label{ga1}
&&\Omega=\sqrt{2R}\,[\,{\cal E}(\epsilon^2)-(1-\epsilon^2){\cal K}(\epsilon^2)\,],~~~~~0<\epsilon<1,~~~~~~\\
&&\Omega=\sqrt{2R}\,\,\epsilon\,\,{\cal E}(\epsilon^{-2}),~~~~~~~~~~~~~~~~~~~~~~~~1<\epsilon,
\label{ga2}
\EEA
where $\epsilon=\sqrt{\frac{E+R}{2R}}$, and where ${\cal E}(x)=\int_0^{\pi/2}\d
\xi\,(1-x\sin^2\xi)^{1/2}$ and ${\cal K}(x)=\int_0^{\pi/2}\d
\xi\,(1-x\sin^2\xi)^{-1/2}$ are the elliptic integrals.  The two regimes
(\ref{ga1}) and (\ref{ga2}) correspond, respectively, to the trapped
and free motions. 

As the feedback function we take 
$\varphi=|\cos(\frac{q}{2})|=\sqrt{1-\sin^2(\frac{q}{2})}$, which is
invariant with respect to $p\to-p$ and $q\to q+2\pi$.  The adiabatic
invariant reads from (\ref{ff})
\BEA
&&I=\frac{\pi}{2}\,\epsilon^2\,\sqrt{\frac{R}{2}},
~~~~~~~~~~~~~~~~~~~~~~~~~~~~0<\epsilon<1,~~\\
&&I=\epsilon\,\sqrt{\frac{R}{2}}\, \left[\,
\sqrt{1-\frac{1}{\epsilon^2}}+\epsilon\,
\arcsin(\frac{1}{\epsilon})\, \right], ~~ 1<\epsilon.~~~~~~
\label{ga4} \EEA The two regimes have the same meaning as above. The
behavior of $I$ and $\Omega$ is presented in Fig.~\ref{fig_1}.
Particles with the energy above the initial separatrix $E=R_{\rm
i}$, but below the final separatrix $E=R_{\rm f}$ get trapped.  In
the non-feedback case the energy $E$ responds to the amplitude
increase in a non-linear way: the low energy motion decreases its
energy, as intuitively expected for almost harmonic oscillations,
while the energies around the separatrix increase. In contrast, 
the feedback {\it linearizes} the
energy response: now all energies decrease. This is because due to 
$\dot{R}\propto|\cos(\frac{q}{2})|$, the amplitude does not move
when the particle is around the unstable fixed points $q=\pm\pi$, which are
responsible for the strong nonlinearity.

We shall now combine the two scenarios (with and without feedback)
so as to get trapping via a cyclic change of $R$.  
Apply first the feedback control
changing $R$ from $R_{\rm i}=1$ to $R_{\rm f}=2$; see
Fig.~\ref{fig_1}. Then apply the
non-feedback control bringing $R$ back to its initial value $1$. It
is seen from Fig.~\ref{fig_1} and (\ref{ga1}--\ref{ga4}) that all
energies $E<E_{\rm m}=1.249$ get trapped after the cyclic process.
Here $E_{\rm m}$ is defined via solving two coupled equations:
$I(E_{\rm m}, R_{\rm i})=I(E^*, R_{\rm f})$ and $\Omega(E=R_{\rm i},
R_{\rm i})=\Omega(E^*, R_{\rm f})$, i.e., in the feedback part of
the process $E_{\rm m}\to E^*=1.129$, while in the non-feedback part
$E^*$ goes precisely to the separatrix $E=R_{\rm i}$. 
\begin{table}[bhb]
\caption{
Numerical illustration of the cyclic trapping method.  The energies are
obtained from solving numerically the oscillator equation of motion
$\ddot{q}+R_t\sin q=0$ and the feedback equation
$\tau_R\dot{R}_t=|\cos\frac{q}{2}|$ in the time-interval from $t=0$ till
$t_1=2\times 10^3$. Here $R_{\rm i}=R(0)=1$ and $\tau_R=10^{-3}$.  Then
a non-feedback process was applied via equation of motion
$\ddot{q}+[R(t_1)-\frac{t-t_1}{\tau_R}]\sin q=0$ till the final time
$t_{\rm f}=t_1+\tau_R [R(t_1)-R_{\rm i}]$, so that $R(t_{\rm f})=R_{\rm
i}=1$.  The adiabatic invariant (\ref{ga4}) is conserved:
$\frac{I_{\rm f}-I_{\rm
i}}{I_{\rm i}}\simeq 2\times 10^{-4}$. 
Expectedly, its dependence on the initial conditions $(q(0),
p(0))$ for a given $E(0)$ is very weak. 
The initial energies $E_{\rm i}=E(0)$
are above the separatrix $E=1$ (free motion), while the final ones are
below it (trapping). 
}
\begin{tabular}{|c||c|c|c|c|c|c|}
\hline 
\,\,$E_{\rm i}$\,\,  & 1.26       & 1.22      & 1.2       & 1.15       & 1.1       & 1.     \\
\hline  
\,\, $E_{\rm f}$\,\, & \,0.9976\, &\,0.9791\, &\,0.9709\, &\,0.9423\,  &\,0.9111\, &\,0.8429\,  \\
\hline
\end{tabular}
\label{tab1}
\end{table}
These arguments are qualitative, since in the feedback case $R_{\rm f}$
is not given {\it a priori}, but comes out from the common solution of
(\ref{hek2}, \ref{de}) and can depend on $E_{\rm i}$.  However,
the numerical results presented and explained in Table I confirm
the arguments. 

In conclusion, we presented an adiabatic feedback method for controlling
classical Hamiltonian systems.  The method is based on an
adiabatic invariant. This requires ergodicity (time average is
equal to the microcanonical one) of certain observables and makes explicit
the change of the microcanonical entropy due to control. We
illustrated the method via two basic models of the non-linear science:
Fermi's accelerator
and the non-linear
oscillator. For the Fermi accelerator|which was intensively studied as
model of plasma heating \cite{zas,liebe_g}|the adiabatic feedback method
offers efficient schemes of heating that, in particular, may lead to
entropy decrease.  For the non-linear oscillator (wave-particle
interaction) transitions are possible between ergodic and non-ergodic
regimes of motion \cite{zas}. They correspond to the particle
(de)trapping by the wave.  We designed schemes for particle trapping via
cyclic change of the wave amplitude, and noted how the feeedback changes
qualitatively the response of this non-linear model.  It was seen that
when controlling an ergodic $n$-particle system ($n\gg 1$) via
monitoring few of its coordinates, the entropy decrease $\Delta S$ due
to feedback is $\Delta S\lesssim \ln n$, while the entropy itself is
$S\sim n$. This is not unexpected, since, e.g., the entropy difference between
an organism and a chemical substance of the same weight is also much
smaller than each of those entropies \cite{blum}.  If, however, there is
an efficiently controlled ($\Delta S\sim S$) macroscopic system, we
foresee it to be non-ergodic, so that the adiabatic invariance is absent. 

A.E. A. and D.B. S. were supported by CRDF grant ARP2-2647-YE-05.
K.G. P. was supported by ISTC grant A-820.


\begin{thebibliography}{99}
\bibitem{com1}For a stochastic motion of the piston it may be possible
to increase indefinitely the average energy, but then fluctuations of
energy also grow considerably \cite{zas}.
\end{thebibliography}
\end{document}